\documentclass[aps,pra, twocolumn,superscriptaddress, showpacs]{revtex4-1}
\usepackage{graphicx}
\usepackage{amsmath}
\usepackage{amssymb,times}
\usepackage{color}

\newcommand{\Spec}{{\rm Spec}}
\newcommand{\Sraw}{{S}^{{\rm raw}}}

\newcommand{\ket}[1]{\left| #1 \right\rangle}
\newcommand{\bra}[1]{\left\langle #1 \right|}

\newcommand{\proj}[1]{| #1\rangle\!\langle #1 |}
\newcommand{\Tr}{\mathrm{Tr}}

\newcommand{\mX}{\mathbb{X}}
\newcommand{\mY}{\mathbb{Y}}
\newcommand{\mZ}{\mathbb{Z}}
\newcommand{\mI}{\mathbb{I}}
\newcommand{\mS}{\mathbb{S}}

\newcommand{\be}{\begin{equation}}
\newcommand{\ee}{\end{equation}}
\newcommand{\bea}{\begin{eqnarray}}
\newcommand{\eea}{\end{eqnarray}}

\newcommand{\bean}{\begin{eqnarray*}}
\newcommand{\eean}{\end{eqnarray*}}

\date{\today}
\begin{document}
\title{Macroscopic instructions vs microscopic operations in quantum circuits}
\author{A.~Veitia}
\affiliation{Department of Physics and Center for Optical, Molecular and Quantum Sciences,
University of Oregon, Eugene, OR 97403}
\author{M.P.~da Silva}
\affiliation{Rigetti Computing, 775 Heinz Avenue, Berkeley, CA 94710}
\author{R.~Blume-Kohout}
\affiliation{Sandia National Laboratories, Albuquerque, New Mexico 87185}
\author{S.J.~van Enk}
\affiliation{Department of Physics and Center for Optical, Molecular and Quantum Sciences,
University of Oregon, Eugene, OR 97403}
\begin{abstract}
 In many experiments on microscopic quantum systems, it is implicitly assumed that when a macroscopic procedure or ``instruction" is repeated many times -- perhaps in different contexts -- each application results in the same microscopic quantum operation. But in practice, the microscopic effect of a single macroscopic instruction can easily depend on its context. If undetected, this can lead to unexpected behavior and unreliable results. Here, we design and analyze several tests to detect context-dependence. They are based on invariants of matrix products, and while they can be as data intensive as quantum process tomography, they do not require tomographic reconstruction, and are insensitive to imperfect knowledge about the experiments. We also construct a measure of how unitary (reversible) an operation is, and show how to estimate the volume of physical states accessible by a quantum operation.
\end{abstract}
\maketitle

\begin{figure}[h!]
 \begin{center}
\includegraphics[width=2.75in]{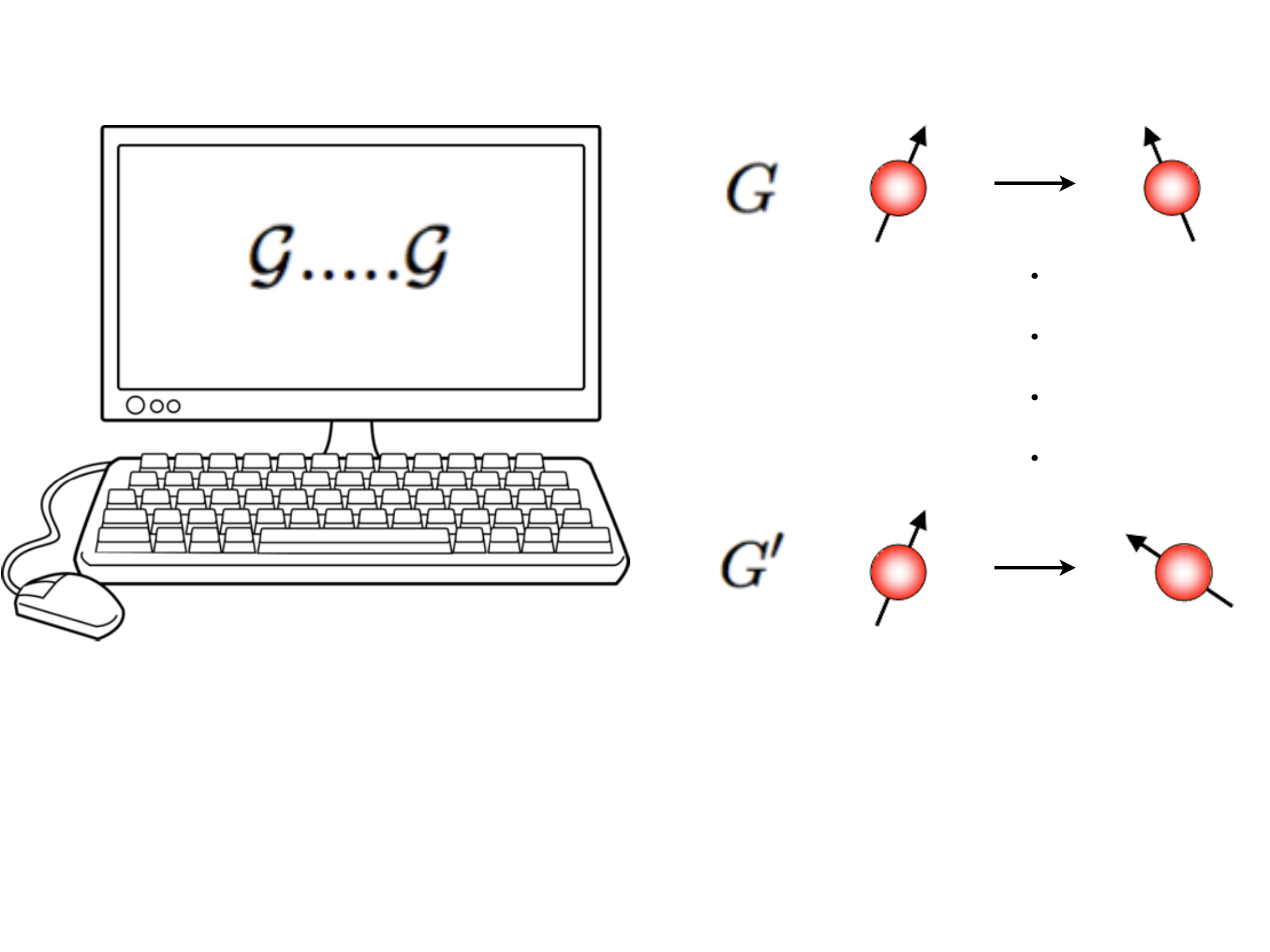}
\vspace{-5em}\caption{If the computer instructions for two particular operations in a sequence are the same (both are ${\cal G}$), are the actual applied quantum operations, $G$ and $G'$, the same, too? Or is there some {\em context} dependence (see main text for a definition) that makes the microscopic operations different?}
\label{fig.context}
\end{center}
\end{figure}

In many modern physics experiments, a fixed and repeatable ``macroscopic" procedure is performed with the intent of effecting a specific action on a microscopic quantum system.  For example, the spin of a single NV center in diamond \cite{golter2014} can be rotated by applying a precise combination of laser fields with specific durations, intensities, and polarizations.  Or the hyperfine ground state of a single trapped ion can be prepared, and then rotated to a different hyperfine state, again using precise control of lasers \cite{monroe1995}.\\
{\indent}More generally, many quantum computing experiments involve applying thousands of quantum operations (called \emph{gates} in that context; we use both terms interchangeably) to one or two qubits at a time. These experiments share a common description, whether the qubits are trapped ions, superconducting circuits in microwave cavities, or individual photons in waveguides, in terms of \emph{quantum circuits} that constitute sequences of instructions  describing  gates  to  be  applied  to  qubits.  Ideally, these gates would be in 1:1 correspondence with specific unitary operations applied to qubits.  But in real-world experiments on imperfect qubits, this is not quite true. An experimentalist (or the computer controlling her experiment) reads the list of gate instructions, and physically implements each one. No two implementations of the same gate -- even immediately successive ones -- are quite perfectly identical (see Fig.~\ref{fig.context}). We say that the real operations depend on their  {\em context}.  A gate's context includes all the external variables that influence it -- e.g., temperature (for a nice example relating to NV centers in diamond, see Ref.~\cite{fu2009}), stray magnetic fields, the local charge environment, time of day, and many others. These variables may be classical or quantum.\\
{\indent}Context-dependence is often neglected, for two good reasons. First, any variables that do  not  change over the course of an experiment (which usually includes multiple repetitions of multiple circuits), or do  not  influence the quantum operation, can and will be ignored. Second, a context variable that varies randomly and is identically and independently distributed ({\em iid}) at every application of a gate can also be eliminated, by simply replacing the ideal operation with its average  over the {\em iid} random values of the context. This is very common, and leads to replacing unitary operations with non-unitary completely positive trace preserving (CPTP) maps \cite{choi1975completely,jamiolkowski1972linear,Gorini1976,Lindblad1976,ruskai2002analysis,nielsen2010quantum} described  by process matrices. It is widely appreciated that operations on real quantum processors are not unitary, and that CPTP maps are a better model \cite{o2003demonstration,fedorov2012implementation,childs2001realization,monz2009realization,rodionov2014compressed,dicarlo2009demonstration}. It is somewhat less widely appreciated  that real operations may also fail to be described by CPTP maps (see, e.g., \cite{Rivas2014}). This can happen whenever the implementation of a gate instruction depends on a context that  is  not  {\em iid}.  Our goal in this paper is to define some easy-to-perform tests for this phenomenon, which we will refer to generically as context-dependence (note that we do not  intend this to include standard, uninteresting {\em iid} context-dependence where each gate can be modeled by a CPTP map).\\
{\indent}It is useful to divide contexts into two categories:  those that are extrinsic  to the quantum circuit, and those that are intrinsic  to it. This distinction is useful when a circuit is repeated many times and the repetitions are treated as exchangeable, as is done in characterization and benchmarking experiments like gate-set-tomography (GST) \cite{merkel2013,blume2013,greenbaum2015} and randomized benchmarking \cite{Emerson2005,Knill2008,Magesan2011,fogarty2015}. In this situation, a context that influences a gate operation is extrinsic if it is not correlated with that gate's role or location in 
the circuit. More precisely: Let $\mathcal{G}$  be a gate instruction, $q$ a variable describing a context that influences the effect of $G$, and $C(\mathcal{G})$ be a complete description of $\mathcal{G}$'s role in a particular circuit, including the circuit itself and where $\mathcal{G}$ appears in it. Now, $q$ is extrinsic if its marginal distribution is identical at every possible circuit location. Extrinsic contexts include time, electromagnetic fields, and the state of a spin or photon bath. Intrinsic contexts are variables that do depend on circuit location, such as the number of gates already performed in the circuit (if gates cause heating), and the identity of the immediately preceding gate (if ring-down causes pulses to overlap, or if the preceding gate disturbed the environment). The tests we construct here are capable of detecting both kinds of context-dependence, but they do not generally distinguish between them.\\
{\indent}To test for context dependence, we start by defining what can happen if it is \emph{not} present. A context-\emph{independent} gate always does the same thing to the underlying microscopic quantum system, which we assume can be described using a $d$-dimensional Hilbert space (for known $d$), so its state is a $d\times d$ density matrix $\rho$ in the $d^2$-dimensional space of Hermitian matrices. Context-independent gates can be represented by fixed $d^2 \times d^2$ \emph{process matrices} acting linearly on $\rho$. These combine associatively; for operations represented by matrices $A$ and $B$, ``First apply $B$, then $A$" is represented by the matrix $AB$.\\
{\indent}We want tests for context-dependence that are independent of (1) what the gates do, (2) whether we know what they do, and (3) ``SPAM" (state preparation and measurement) errors. To achieve this robustness, we test directly for \emph{violation of associativity}, using three simple facts. First:  for any sequence of $n\times n$ matrices $\{A_k\}^m_{k=1}$, the spectrum of their product is invariant under cyclic permutations of the list. Second:  the determinant of that product is invariant under \emph{any} permutation. Third:  for specific sets of matrices, the determinant of their product decays exponentially with $m$ (the length of the list). Note that a different approach for detecting context-dependence has been recently proposed in Ref.~\cite{rudinger2019probing}, which consists in carefully examining how the statistics of the measurements outcomes associated with a set of operations (circuit) are affected by changes in the circuit's context (such as acting on neighboring qubits). 

There is no general theory for modeling arbitrary context dependence -- but we do not need one. We need only detect deviations from the null hypothesis of well-behaved context-independent gates. Under \emph{that} assumption, any sequence of gates $S$ transforms the system's state $\rho$ by a $d^2\times d^2$ matrix ${S}$ with elements 
\be
\label{PL}
{S}_{nm}=\frac{1}{d}\Tr\big[P_n S(P_m)\big],
\ee
where $\{P_n\}_{n=1}^{d^2}$ is a Hermitian and orthogonal \footnote{We choose the normalization $\Tr(P_{n}P_{m})=d\delta_{nm}$} basis of operators on $\mathcal{H}_d$. If we could directly measure ${S}$ for any sequence of operations $\mathcal{S}$, testing for context dependence would be trivial.  We would just figure out the process matrix for each macroscopic elementary gate $G_i$, pick some sequences of those gates, measure \emph{their} process matrices, and check associativity (e.g., does $A\cdot B$ represent $A\circ B$?). This direct and unambiguous reconstruction of ${S}$ was the goal of \emph{quantum process tomography}~\cite{PoyatosCirac1997,ChuangNielsen1997}, but it runs afoul of the problem that we generally can't inject perfect known matrices $P_n$ into the system and measure their expectations. It would be sufficient to inject perfectly known states $\rho_i$ and measurements $\Pi_k$, but in practice states and measurements are implemented using the same unknown (and unreliable) gates that we want to characterize, and this makes process tomography unreliable~\cite{Schulman2012,blume2013}. But, remarkably, we can use unreliable process tomography estimates to construct \emph{reliable} witnesses for context dependence, based on the spectral properties listed above. \\
{\indent}To do process tomography on a sequence of instruction $\mathcal{S}$, we first construct a probability table $\mathcal{P}_{k|i}(\mathcal{\mathcal{S}})$ in three experimental steps:  (i) create $d^2$ linearly independent states; (ii) apply $S$; and (iii) measure $d^2$ linearly independent POVM effects (see note \footnote{Any quantum measurement can be represented by a \emph{positive operator-valued measure} (POVM) comprising a set $\{E_{m}\}$ of positive $d\times d$ operators called \emph{effects} that sum to $I_{d}$. In step (iii), it's not important how many distinct POVMs are needed to realize the $d^2$ distinct effects.}). This procedure defines $d^4$ distinct events -- each of the form ``We prepared $\rho_i$, did $S$, and then observed outcome $\Pi_k$" -- whose probabilities can each be estimated by repeating an experiment $N_{s}\gg1$ times and observing how many times ($n$) the event in question happened. Between $d^2$ and $d^4$ distinct experiments are required (depending on how many outcomes each POVM measurement has), and these should be performed in random or interleaved fashion to average out the effects of simple drift \cite{vanenk2013}.\\
{\indent}\emph{Ideally}, the first step would prepare known and linearly independent pure states $\rho_i = \proj{\phi_i}$, and the last would perform projective measurements with linearly independent effects $\Pi_k = \proj{\psi_k}$ that span the vector space of Hermitian matrices as uniformly as possible (e.g., mutually unbiased bases would do nicely). Then, from the observed frequencies, we would construct a ``raw" process tomographic estimate 
\be\label{Sraw}
\Sraw=({\cal{B}}^{-1})^T
{\mathcal{P}}(\mathcal{S}){{\cal{C}}}^{-1},
\ee
where the entries of the $d^{2}\times d^{2}$ matrices $\mathcal{P}(\mathcal{S})$, ${\cal{B}}$ and ${\cal{C}}$ are
${\mathcal{P}}_{k|i}(\mathcal{S})$, $B_{nk}=\Tr(P_n\Pi_k)/\sqrt{d}$, and 
$C_{mi}=\Tr(P_m\rho_i)/\sqrt{d}$, respectively. In the absence of SPAM errors we would simply have ${S}^{\text{raw}}={S}$, up to sampling errors.\\
{\indent}{\em{In practice}}, input states are prepared by applying a set of gates  $\{G^{\text{in}}_{i}\}_{i=1}^{d^{2}}$ to the initial state of the system $\rho_{0}$. Then, after $S$, a set of gates $\{G^{\text{out}}_{k}\}_{k=1}^{d^{2}}$ can be used to rotate the measurement axes, before measuring a fixed POVM effect $M_{0}$.  Assuming context-independence,  we have $\mathcal{P}_{k|i}(\mathcal{S})=\Tr(M_{0}G^{\text{out}}_{k}\circ S \circ G^{\text{in}}_{i}(\rho_{0}))$. In order to connect $\Sraw$ to the matrix
${S}$ representing the actual process $S$ we applied,
we define two unknown linear maps $E_{{\rm in}}$ and $E_{{\rm out}}$ such that
\bea\label{Ein}
E_{{\rm in}}(\rho_i)&=&G_i^{{\rm in}}(\rho_0),\nonumber\\
E_{{\rm out}}(\Pi_k)&=&G_k^{\dagger {\rm out}}(M_0),\label{Einout}
\eea
\footnote{The adjoint of the map here is defined w.r.t. the Hilbert-Schmidt inner product.}. These two maps need not be physical (they may be non-trace-preserving, for instance); they merely provide a mathematical description of SPAM errors. Note that these maps exist provided the sets of preparations and measurements are not {\em overcomplete}. Moreover, Eqs.~(\ref{Einout})  are guaranteed to have a unique solution when both $\{\rho_i\}_{i=1}^{d^{2}}$ and $\{\Pi_{k}\}_{k=1}^{d^{2}}$ span the space of linear operators on ${\cal H}_d$. Now,  $\Sraw$ is related to $S$ by 
{\be\label{Soraw}
\Sraw={E}^T_{{\rm out}}{S}{E}_{{\rm in}},
\ee
where the real-valued matrices ${E}_{{\rm in},{\rm out}}$ represent the corresponding (linear) maps $E_{{\rm in},{\rm out}}$. This relationship holds as long as all the operations are context-independent.\\
{\indent}First, consider a length-$m$ sequence of instructions $\mathcal{S}_{1} = \mathcal{G}_{m}\circ \mathcal{G}_{m-1}\circ \ldots \circ \mathcal{G}_{1}$ and any permutation $\mathcal{S}_{\sigma} := \mathcal{G}_{\sigma(m)}\circ \mathcal{G}_{\sigma(m-1)}\circ \ldots \circ \mathcal{G}_{\sigma(1)}$ of it. Notice now that if each instruction $\mathcal{G}_{i},$ for $i=1\ldots m,$ gives rise to a context-independent gate $G_{i},$ we learn from Eq.~(\ref{Soraw}) that
\begin{eqnarray}
\label{test1}
\det (\Sraw_\sigma)&=&\det( {E}_{{\rm out}}{E}_{{\rm in}}) \prod_{i=1}^m \det(G_{\sigma(i)})\nonumber \\
&=& \det( {E}_{{\rm out}}{E}_{{\rm in}})\prod_{i=1}^m \det(G_{i})=\det(\Sraw_{1}), 
\end{eqnarray}
for any permutation $\sigma.$ This equality implies that any statistically significant variation of the quantity $\det(\Sraw_{\sigma})$ (with $\sigma$) is incompatible with a model wherein the gates $\{G_{i}\}_{i=1}^{m}$ are context-independent. In addition, note that SPAM errors are explicitly included here (via the matrices $E_{\text{in}}$ and $E_{\text{out}}$) and  cannot, therefore, cause a false alarm. However, this test will miss some forms of context-dependence -- e.g. when a gate is a unitary operation whose identity depends on the previous gate. We will refer to this context-independence test, based on Eq.~(\ref{test1}), as the permutational determinant (PD) test.\\
{\indent}More tests, which make use of the entire spectrum, not just the determinant, can be constructed by  adding a short \emph{reference sequence} $S_{0},$ which is to be considered as part of SPAM and included in {\em all} experiments. In the absence of context-dependence, our tests will not depend on the choice of the sequence $S_{0},$ which can then be treated as an extra error in the gates $\{G_{i}^{\text{in}}\}_{i=1}^{d^{2}},$ or  $\{G_{k}^{\text{out}}\}_{k=1}^{d^{2}}$ \footnote{The simplest choice of the reference sequence $S_{0}$ is the identity matrix $I_{d^{2}},$ corresponding to the instruction to do nothing between the gates $\{G_{i}^{\text{in}}\}$  and $\{G_{k}^{\text{out}}\}.$ We could, however, run all our tests for different reference sequences. The results ought to be the same irrespective of the (short!) reference sequence. This provides a meta test of our tests.}. We can then apply Eq.~(\ref{Soraw}) to the raw data from {\em just} the short experiment (the type of data used for SPAM tomography \cite{stark2014,jackson2015}) and write 
\be\label{Sraw0}
\Sraw_0={E}_{{\rm out}}^T{E}_{{\rm in}}.
\ee
Since the spectrum of a matrix is invariant under similarity transformations, we find combining equations (\ref{Soraw}) and (\ref{Sraw0}) that
\begin{eqnarray} 
\label{Spec}
\Spec(S) &=&  \Spec[E_{\rm{out}}^{T} S (E_{\rm{out}}^{T})^{-1}] \nonumber \\
                     &=&\Spec[(E_{\rm{out}}^{T} S E_{\rm in}) (E_{\rm out}^{T} E_{\rm in})^{-1}] \nonumber \\
		    &=& \Spec[\Sraw (\Sraw_0)^{-1}], 
\end{eqnarray} 
which allows us to extract the spectrum of any sequence $S$ from raw tomographic data. Now, in a similar fashion to the test Eq.~(\ref{test1}), we can consider the $m$ cyclic permutations $\sigma'$ of a sequence of $m$ instruction and examine, via Eq.~(\ref{Spec}), whether $\Spec({S}_{\sigma'})$ remains invariant (as it should in the absence of context-dependence). Equivalently, by virtue of the Cayley-Hamilton theorem and Eq.~(\ref{Spec}), we can phrase this \emph{cycle-test} in terms of the invariance of the traces 
\be\label{test2}
{\cal F}^{(r)}_{\sigma'}:=\frac{1}{d^2}\Tr({S}^r_{\sigma'})=\frac{1}{d^2}\Tr([\Sraw_{\sigma'} (\Sraw_0)^{-1}]^r),
\ee
for $r=1\ldots d^2.$ Note that ${\cal F}^{(r)}=\frac{1}{d^2}\Tr({S}^r)$ is just the process fidelity \cite{Horodecki1999,Nielsen2002simple} of the operation ${S}^r = S\circ \ldots \circ S$ ($r$ times), with respect to the identity.

It is worth mentioning that the above permutational invariants can be expressed directly in terms of the  $d^{2}\times d^{2}$ probably matrices ${\mathcal{P}}(\mathcal{S}).$ More precisely, making use of equations (\ref{Sraw}), (\ref{test1}) and (\ref{test2}) one can easily show that context-independence implies that {(\bf{i})} $\det ({\cal P}(\cal S_\sigma))$ is  invariant under permutations $\sigma$ and {(\bf{ii})} $\Spec ({\cal P}(\mathcal{S}_{\sigma'}){\mathcal P}_0^{-1})$ is invariant under cyclic permutations $\sigma'$, where ${\mathcal P}_0=\mathcal{P}(\mathcal{S}_{0})$ is the probability matrix obtained from the short experiment (see discussion around Eq.~(\ref{Sraw0})). Formulating these tests directly in terms of the data, removes any need to estimate gates or SPAM. It also demonstrates that they are gauge-invariant, in the terminology of GST \cite{merkel2013,blume2013,greenbaum2015,rudnicki2017}.

The permutational tests are only useful for sequences involving at least two different gates (one of which could be the idle gate), so context-dependence cannot be isolated to a single gate \footnote{The data obtained from permutation tests are still useful for model selection purposes \cite{model,schwarz2013} in any case.}.
Since tests that target individual gates are useful, for debugging purposes, here is a test that targets individual gates. 

If a gate $G$ is repeated $m$ times, then if all the $\mathcal{G}s$ (see Fig.~\ref{fig.context}) implement the same process matrix ${G}$, then 
\be\label{test3}
L_m=m \log|\det({G})|+\log|\det({E}_{{\rm out}}{E}_{{\rm in}})|,
\ee
where $L_m:=\log|\det(\Sraw_m)|.$ Varying the sequence length $m$ yields the following test: if $L_m$ does not depend linearly on $m$ \footnote{The exponential decay may remind one of Randomized Benchmarking \cite{Emerson2005,Knill2008,Magesan2011,fogarty2015}, but here we obtain this without the need to randomize over different (Clifford) gates. }, then either the gate $G$ and/or SPAM operations must depend on their context. We will refer to this test the iterative determinant (ID) test. If $L_m$ does depend linearly on $m$ (within error bars), then its slope yields the gate-specific quantity $\log|\det({G})|$, which is intrinsically interesting for at least three reasons. First, $|\det({G})|=1$ implies the gate is unitary \cite{Wolf2008}. Deviations from unitarity can be quantified through a measure of unitarity  proposed in \cite{wallman2015}:
\be
u(G)=\frac{1}{d^2-1}\Tr(W^TW),
\ee
where $W$ is called the unital part of ${G}$. We can use this definition to find a lower bound on $u(G)$ in terms of 
$|\det({G})|$ (the proof of this bound is given in an accompanying paper \cite{veitia2018testing})
\be
u(G)\geq |\det({G})|^{\frac{2}{d^2-1}}.
\ee
Now, the measure $u(G)$ has the inconvenient property (noted explicitly in \cite{wallman2015}) that
for two gates $G_1$ and $G_2$ we may have
$u(G_2\circ G_1)>u(G_1)$.
The determinant, on the other hand, has the property
$|\det({G}_2{G}_1)|=|\det({G}_1)\det({G}_2)|\leq
|\det({G}_1)|$ {\em and} it is gauge independent, and so it is natural to use our bound to define a new measure of unitarity as
\be
{u'}(G)= |\det({G})|^{\frac{2}{d^2-1}}.
\ee

Second, the determinant of a  process $S$,  $|\det({S})|$, corresponds to the ``volume of accessible states'' of that process~\cite{lorenzo}. If this volume increases in time, $S$ must violate CP divisibility \cite{lorenzo}. If the gates forming $S$ are {\em not} context-dependent, then we can use Eqs.~(\ref{Sraw})--(\ref{Sraw0}) to express this quantity as
\be
|\det({S})|=\left|\frac{\det ({\cal P}(\mathcal{S})}{\det(\mathcal{P}_{0})}\right|,
\ee
and so the accessible volume of the process $S$ is given by a ratio of volumes in probability space. 

Third, the decay of the determinant of a process $S$ is not affected by Hamiltonian evolution, only by decoherence. Specifically, consider the Lindblad master equation 
\be\label{detLind}
\frac{d\rho_{t}}{dt}=-i[H_t,\rho_{t}]+{D}(\rho_{t}),
\ee 
where $H_{t}$ is the system's Hamiltonian while the linear map $D$ describes dissipative effects (see e.g., \cite{wiseman2009quantum}). Let $S_{t}$ be the dynamical map describing the time evolution of the system, i.e., 
$\rho_{0} \rightarrow \rho_{t}=S_{t}(\rho_{0})$, then making use of the representation Eq.~(\ref{PL}) one can prove that \cite{Wolf2008,hall2014,veitia2018testing}
\be
\label{slope}
 \frac{d}{dt}\log\det({S}_t)=\Tr({D}),
\ee
independently of the Hamiltonian $H_{t}.$\\
{\indent}To illustrate how our tests work, we consider and analyze an example where gates on one qubit ($A$) are made context-dependent by an ``unwanted" coupling to another hidden (but persistent) qubit ($B$). To simulate this example numerically, we choose specific parameters: \\
 \noindent {\bf (i)} The four {\em ideal} in/out gates are 
\be
\{G^{{\rm ideal}}\}=\{I,X_{\pi},Y_{-\pi/2},X_{-\pi/2}\}
\ee
which are to be used when calculating
$\Sraw$ (see the discussion around Eq.~(\ref{Sraw})). That is, the ideal states prepared for qubit $A$ for $i=1\ldots 4$ are
$\rho_i=\{G^{{\rm ideal}}\}
\proj{0}$, and the four ideal measurements performed would be 
$\Pi_k=\{G^{{\rm ideal}}\}^\dagger\proj{1}$. \\
\noindent {\bf (ii)}
The qubit-qubit coupling is of the Ising form
\be
V=\frac{J}{2}\sigma_z^A\otimes \sigma_z^B.
\ee
{\bf (iii)} Both qubits experience energy relaxation, thermal excitation, and dephasing, at rates $\gamma_1$, $\gamma_3$ and $\gamma_{\phi}$.\\
\noindent {\bf (iv)}
 Errors in {\em any} gate (including $G_i^{{\rm in (out)}}$ acting on qubit A) are modeled by using the following matrix to represent the noisy gate (which also includes an action on qubit $B$)
\be\label{hG}
\mathbb{G}=\exp({{\cal J}}_G+t_g{V}+t_g{D}),
\ee
where $t_g$ sets the gate duration and ${D}$ generates the decoherence of (iii) above, on {\em both} qubits.  The matrix ${\mathcal J}_G$ generates the ideal gate $G$ on qubit $A$. More precisely, to implement the noisy gate $\mX_\theta$, we choose ${\mathcal J}_{X_\theta}$ to be the matrix representation Eq.~(\ref{PL}) of the map $-i\tfrac{1}{2}\theta [X,.]\otimes I$ that appears in the (two-qubit) Lindblad equation, whereas the noisy idle gate $\mathbb{I}$ is modeled by choosing ${\mathcal{J}}_{I}=0$.\\
\noindent {\bf (v)} An additional state-preparation error is included by applying the operator (\ref{hG})
to the noisy state
$\rho_0'=(p\proj{0}+(1-p)\proj{1})^{\otimes 2}$ (so, qubit A does {\em not} start in $\ket{0}$).
We choose $p=\gamma_1/(\gamma_1+\gamma_3)$ so that the initial state is stationary under pure decoherence. \\
\noindent {\bf (vi)} Measurement errors on $A$ are modeled by replacing the ideal gates by the noisy versions (\ref{hG}) {\em and} by introducing an efficiency $\eta<1$, so that $M_{0}=\eta \ket{1}\bra{1}$. (B is not measured;  it is traced out in the end.)

Fig.~2 shows simulations of our permutational tests. Starting from the sequence
$\mS_1=\mI^{250} \mX_{\pi}^{250}$ (of length $m = 500$) we consider the permutations $\mS_2=\mI^{249} \mX_{\pi}^{249} (\mX_{\pi}\mI) $,
$\mS_3=\mI^{248}\mX_{\pi}^{248}(\mX_{\pi}\mI) (\mX_{\pi}\mI)$ etc., until $\mS_{251}=(\mX_{\pi}\mI)^{250}$. We compute the log of the left-hand side of (\ref{test1}), $L_k=\log|\det(\Sraw_k)|$, for the sequences $\mS_k$ as a function of $k$ for different values of the parameter $\varphi:=Jt_g,$  which sets the strength of the interaction between the qubit and the memory, and thus the magnitude of the context-dependence effects. We chose $p = 0.92, \eta= 0.95, \gamma_1^{-1}
= 60\mu s, \gamma_{\phi} = \gamma_1/2$ and $t_g = 20$ns. For $\varphi=0$ the gates are noisy but not context-dependent and $L_k$ is constant, but for nonzero values of $\varphi$ we observe the non-invariance of $L_k$ under permutations (even for high fidelity gates with small values of $\varphi$ of order $10^{-3}$). The second permutational test  (Fig.~2(b)) is based on Eq.~(\ref{test2}) with $r=2$. We plot the fidelity 
${\cal F}^{(2)}=\Tr({S}^2)/4$ for the cyclic permutations $\mS_1=\mX_{\pi}\mI^{500}$,
$\mS_2=\mI\mX_{\pi}\mI^{499},$ $\mS_3=\mI^2\mX_{\pi}\mI^{498}$ etc., and we infer that the idle gate is context-dependent from the fact that the fidelity ${\cal F}^{(2)}$ is  not constant for $\varphi\neq 0$.\\
\begin{figure}[t!]
\includegraphics[width=3.4in]{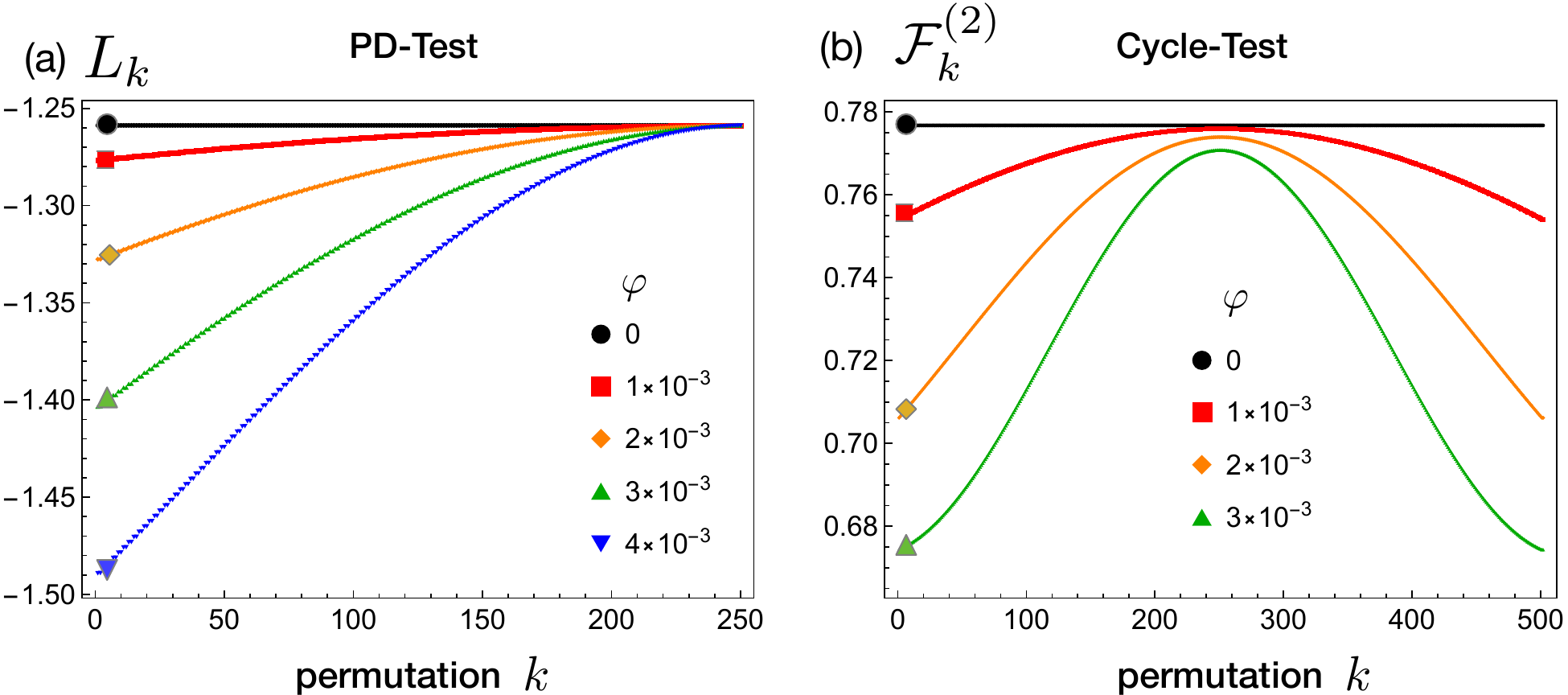}
\caption{Detection of context-dependence via the permutational tests (\ref{test1}) and (\ref{test2}) for various values of $\varphi =J t_{g}.$ The quantities plotted, $L_k=\log|\det({S}_k^{{\rm raw}})|$ and ${\cal F}_k^{(2)}$, are independent of the permutation, labeled by  $k$, of a sequence of instructions provided the gates are not context-dependent.
We considered the permutations  $\mS_{k} = \mI^{251-k}\mX_{\pi}^{251-k} (\mX_{\pi}\mI)^{k-1}, k=1,2,\ldots, 251$ for (a) and the cyclic permutations $\mS_{k} = \mI^{k-1} \mX_{\pi} \mI^{251-k}, k=1,2,\ldots, 251$ for (b).}
\end{figure}
\begin{figure}[t]
 \begin{center}
\includegraphics[width=3.4in]{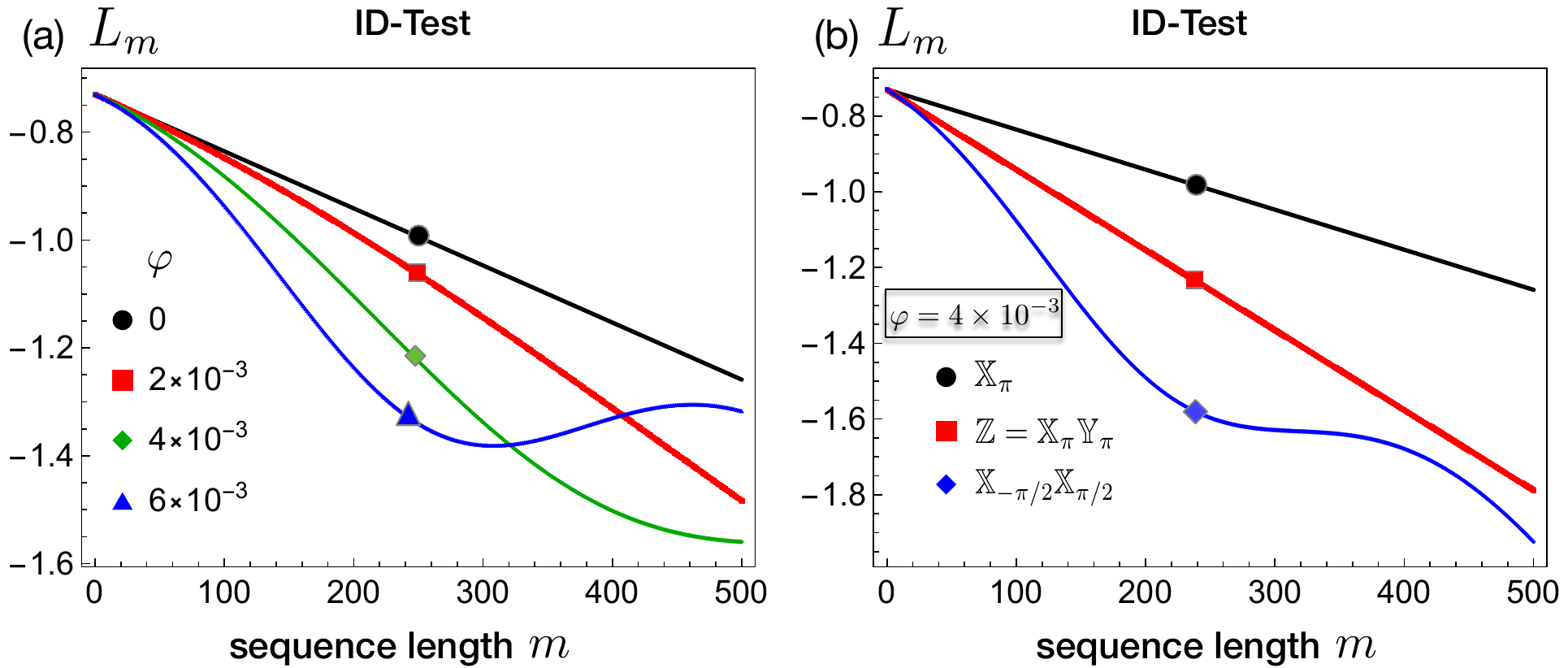}
\caption{Detection of context-dependence via the ID-test (\ref{test3}), for various values of $\varphi = J t_g$.  The quantity $L_m =\log|\det(S^{\text{raw}}_{m})|$ decreases linearly with $m$ for context-independent gates. Panel (a) shows $L_m$  for the noisy idle gate $\mathbb{I}$, applied $m$ times. For larger values $\varphi$ we observe clear deviations from linearity as well as CP-indivisibility via the non-monotonicity of $L_m$ (for the blue $\vartriangle$ curve; see main text). Panel (b) plots $L_m$ for a fixed value of $\varphi$ and three different gates $(\mX_{\pi}, \mZ, \mX_{-\pi/2}\mX_{\pi/2})$ applied $m$ times. Deviations from context-independence are clearly observed in one case (blue $\diamond$ curve).}
\end{center}
\end{figure}
Fig.~3 illustrates tests that apply a single gate $m$ times. In both cases we plot $\log|\det(\Sraw)|$ as a function of $m$, for different values of $\varphi$. Other parameters are as in Fig.~2. Fig.~3(a) shows results for the idle gate, and even for that trivial gate we observe context-dependence. In fact, we also observe (for the largest value of $\varphi$) CP-indivisibility \cite{Wolf2008}: the process for $m=500$ cannot be decomposed into two physical processes of length $m=300$ and one of length $200$, because the determinant cannot increase for a CP-divisible process \cite{Wolf2008}. This conclusion does require assuming that ${E}_{{\rm in,out}}$
are not significantly context dependent~\cite{veitia2018testing}. In Fig.~3(b) we plot the same quantity for three different repetitive sequences of the form $\mX_{\pi}^m, (\mX_{\pi}\mY_{\pi})^m$ and $(\mX_{-\pi/2}\mX_{\pi/2})^m.$ Two of them display no context-dependence but the third (blue $\diamond$ curve) does. Our analytic understanding \cite{veitia2019}  of the former two cases shows there {\em is} in fact no non-Markovianity. The slope of the black curve (cf.~Eq.~(\ref{slope})) is 
$\Tr(D)=-2t_g(\sum_{k=1,3,\phi}\gamma_k$), and for the red one it is twice this slope because 
the duration of the gate $\mZ=\mX_\pi\mY_\pi$ is 2$t_g$.\\
{\indent}We have proposed a family of tests for context-dependence of noise in quantum information experiments, based on invariants of gate sequences. The main attractive features of these tests is that they are gauge invariant as well as robust against SPAM errors, while at the same time not requiring full tomographic reconstruction. The tests also naturally lead to a new measure of unitarity that is monotonic under composition of operations, unlike previously proposed measures. The accompanying publication~\cite{veitia2018testing} examines the effect of statistical fluctuations, which were not addressed here, and suggests a set of tools to test the statistical significance of possible deviations from context-independence. In particular, reference~\cite{veitia2018testing} further exploits the idea of expressing our tests directly in terms of $d^{2}\times d^{2}$ probability matrices (as briefly mentioned in this work), with the purpose of comparing the 
performance of various tomographic schemes, e.g., those based on symmetric informationally complete (SIC) sets~\cite{appleby2014symmetric}. 
\begin{acknowledgements}
A.V., M.P.S. and S.E. were partially supported by ARO under Contract No. W911NF-14-C-0048. M.P.S. performed this work while employed at Raytheon BBN Technologies. Sandia National Laboratories is a multimission laboratory managed and operated by National Technology and Engineering Solutions of Sandia, LLC, a wholly owned subsidiary of Honeywell International, Inc., for the U.S. Department of Energy's National Nuclear Security Administration under contract \text{DE-NA0003525}.
\end{acknowledgements}

\bibliography{black_ops-physlett-v2.bib}

\end{document}